\documentclass[doublecol]{epl2}
\usepackage{bm, amsmath, amssymb}
\usepackage[T1]{fontenc}
\usepackage{textcomp}

\title{Spin Dynamics at the Mott Transition and in the Metallic State
 of the Cs$_{3}$C$_{60}$ Superconducting Phases}
\author{Y.~Ihara \inst{1} \thanks{present address: Department of Physics, Hokkaido university}
\and H.~Alloul \inst{1} \and P.~Wzietek \inst{1} 
\and D.~Pontiroli \inst{2} \and M.~Mazzani \inst{2} \and M.~Ricc{\`{o}} \inst{2}}
\shortauthor{Y.~Ihara \etal}
\date{\today}

\shorttitle{ Spin dynamics and Mott transition in the Cs$_{3}$C$_{60}$ superconducting phases}

\institute{
	\inst{1} Laboratoire de Physique des Solides, Universit\`{e} Psris-Sud 11, CNRS UMR 8502, 91405 Orsay, France\\
	\inst{2} Dipartimento di Fisica, Universit{\`{a}} di Parma - Via G.P.Usberti 7/a, 43100 Parma, Italy
}

\abstract{
We present here $^{13}$C and $^{133}$Cs NMR spin lattice relaxation 
$T_{1}$ data in the A15 and fcc-Cs$_{3}$C$_{60}$ phases for increasing
hydrostatic pressure through the transition at $p_{c}$ from a Mott insulator
to a superconductor. We evidence that for $p\gg p_{c}$ the $(T_{1}T)^{-1}$
data above $T_{c}$ display metallic like Korringa constant values which
match quantitatively previous data taken on other $A_{3}$C$_{60}$ 
compounds. However below the pressure for which $T_{c}$ goes through a
maximum, $(T_{1}T)^{-1}$ is markedly increased with respect to the Korringa
values expected in a simple BCS scenario. This points out the importance of
electronic correlations near the Mott transition.  For $p \gtrsim p_{c}$
singular $T$ dependences of $(T_{1}T)^{-1}$ are detected for $T\gg T_{c}$. 
It will be shown that they can be ascribed to a large variation with
temperature of the Mott transition pressure $p_{c}$ towards a liquid-gas
like critical point, as found at high $T$ for usual Mott transitions. 
}

\pacs{71.30.+h}{Metal-insulator transitions and other electronic transitions}
\pacs{74.70.Wz}{Carbon-based superconductors}
\pacs{74.25.nj}{Nuclear magnetic resonance}

\begin{document}
\maketitle

\section{Introduction}

\begin{figure}[t]
\begin{center}
\includegraphics[width=7.5cm]{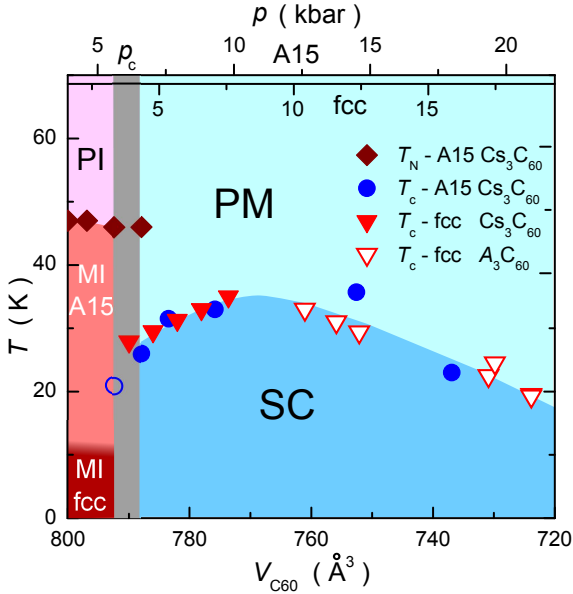}
\end{center}
\caption{(colour on line) Phase diagram representing $T_{c}$ versus the
volume $V_{C_{60}}$ per C$_{60}$ ball, and the Mott transition at $p_{c}$
(hatched gray bar). The magnetic insulating (MI) phase is AF with $T_{N}=47$ K  for
the A15 phase. For the fcc phase a less characterized magnetic state occurs
below $10$ K. The corresponding pressure scales for the two phases are shown on
the upper scale.}
\label{Common phase diagram}
\end{figure}

Among the various systems presenting a transition from a magnetic to a high $%
T_{c}$ superconducting (SC) state, the recently discovered fulleride A15-Cs$%
_{3}$C$_{60}$ phase \cite{takabayashi-Science323} takes a very special
place. Indeed, as in cuprates, the transition observed there is from a Mott
insulating state to a SC state, contrary to Fe pnictides in which SC and
metallic magnetism are contiguous. We have found \cite{ihara-PRL104} that
the pressure controlled transition occurs at different pressures $p_{c}$  of
3.5(5) kbar and 6.5 (5) for the two isomeric structures of Cs$_{3}$C$_{60}$
(A15 and fcc respectively). Below $p_{c}$ the ground magnetic states of the
insulating phases were found distinct for the two structures.  
The A15 is a N\'{e}el  antiferromagnet (AF) while the fcc phase, 
which is not a bipartite structure, displays a more complicated magnetism.
Otherwise the
phase diagrams can be mapped into a unique one, recalled in 
Fig.~\ref{Common phase diagram}, 
when plotted versus $V_{C_{60}}$, the unit volume per C$_{60}$
molecule (as confirmed independently in Ref.~\cite{ganin-Nature466}). 
This established that the interball distance is the driving parameter
for the electronic properties of both compounds. The phase diagram displays
a SC dome, reminiscent of that found in cuprates, with a $T_{c}$ maximum
near the first order phase boundary below which the insulating Mott state
sets in.

But this analogy should be taken cautiously, as in the fullerides the
transition occurs in a three-dimensional (3D) crystal structure with a fixed
number of carriers, while in cuprates it appears in a two-dimensional (2D)
lattice of Cu sites, with hole or electron doping \cite{lee-RMP78}, which is
furthermore a source of dopant induced disorder of prime importance for the
physical properties \cite{alloul-EPL91}. More importantly $A_{3}$C$_{60}$
have been so far considered as standard BCS(-type) phonon driven
superconductors \cite{gunnarsson-RMP69}, with a clear singlet $s$-wave
pairing, distinct from the $d$-wave pairing demonstrated in cuprates 
\cite{tsuei-PRL73} or anticipated in 2D organics \cite{lefebvre-PRL85}

One should notice that such a phase diagram had been anticipated by Capone 
\textit{et al.}, from ground state calculations assuming an electron phonon
mechanism for SC \cite{capone-PRB62, capone-Science296, capone-RMP81}. 
The electronic correlations are of course embedded in this
theoretical treatment in which a low spin $S=1/2$ state is favoured by the
on-ball Jahn-Teller effect for the three electrons transferred on the 
C$_{60}$ balls. 
These authors predicted that pure BCS equations cannot apply
and that the normal state near the Mott transition, where $T_{c}$ increases
with pressure, is a non Fermi liquid exhibiting a moderate increase of the
spin susceptibility or effective mass.

Here we report NMR data on the electronic properties of these compounds
which can be compared to these predictions. We demonstrate that the nuclear
spin lattice relaxation $T_{1}$ data on $^{13}$C and $^{133}$Cs taken near
the Mott transition can by no way be described by a simple pressure induced
modification of the density of states in a non correlated band. On the
contrary magnetic fluctuations concentrate near the Mott transition and only
progressively give place to those resembling a Fermi liquid through the
pressure for which $T_{c}$ is maximum. We give here evidence for a step
increase of $T_{1}^{-1}$ at the critical pressure $p_{c}$, 
a behaviour quite analogous to that seen in
organic compounds displaying a Mott transition \cite{kagawa-NP5}. 
We find that $p_{c}$ increases with $T$ as expected from the 
lattice expansion. 
Furthermore  $p_{c}(T)$ apparently terminates at a critical point 
located near room temperature.

\section{Experimental techniques}

One only expects limited accuracy from $^{13}$C NMR shift measurements in
fullerene compounds in view of the large anisotropy of hyperfine couplings
for the $^{13}$C nucleus \cite{pennington-RMP68}. We could however evidence
that way \cite{ihara-PRL104} the large paramagnetic response in the Mott
state of the Cs$_{3}$C$_{60}$ phases. But, in fullerene compounds,  $T_{1}$
data have been generally more useful \cite{brouet-PRL82, brouet-PRB66}.

Let us recall that $T_{1}$ allows to probe the wave vector $\mathbf{q}$
dependent dynamic spin susceptibilities $\chi (\mathbf{q},\omega )$ of the
electron spins through the general relation 
\begin{equation}  \label{eq:T1-g}
\frac{1}{T_{1}T}= \frac{2k_{B}}{(\gamma_{e} \hbar)^{2}} \sum_{\mathbf{q}} %
\left[ A_{\mathrm{hf}}(\mathbf{q}) \right] ^{2}\mathrm{Im}\left[ \frac{\chi (%
\mathbf{q},\omega )}{\omega } \right],
\end{equation}%
Here the $\mathbf{q}$ dependence of the hyperfine coupling constant 
$A_{\mathrm{hf}}(\mathbf{q})$ 
of the probe nucleus 
is determined by its spatial location with respect to the magnetic sites. 

We have therefore undertaken some $^{13}$C NMR $T_{1}$ measurements on
both A15 and fcc phases, as this allows comparison with the available data
for the other alkali $A_{3}$C$_{60}$ fullerides \cite{pennington-RMP68,
maniwa-JPSJ63}, the  hyperfine coupling being then local, i.e. independent
of the lattice structure. We have selected here two samples described in
Ref.~\cite{ihara-PRL104} with respective C$_{60}$ molecular concentration 
of  A15, fcc and Cs$_{4}$C$_{60}$ phases of 
(A1: $41.7~\%,12~\%,46.5~\%$) and 
(F1: $34~\%,55~\%,11~\%$). 
This allows us avoiding contamination by Cs$_{4}$C$_{60}$ for the
dominant fcc phase sample and by the fcc phase for the dominant A15 phase.
We have also taken advantage of the selectivity permitted by the $^{133}$Cs
NMR \cite{ihara-PRL104}, with three resolved Cs sites (O, T, T'), to study
the $T_{1}$ variations in a large $(p,T)$ range on sample F1. Furthermore,
in this case the absence of quadrupole effects yields exponential recoveries
of the nuclear magnetization.

The pressure experiments on the fcc phase were not performed 
using a classical clamp cell which might yield inaccurate pressure control 
as we are lacking temperature insensitive pressure sensors. 
Here we used an externally controlled system 
in which a pressure generator is connected to the pressure cell through a capillary tubing. 
The pressure is transmitted to the sample space via a pressure multiplier
mounted in the pressure cell. 
We have used isopentane and Flourinert-77 (3M) as 
pressure mediaa for the first and second stage respectively. 
Within this media setup we were able to accurately adjust the 
pressure which is measured at room $T$ while cooling the cell, 
until the complete solidification of Flourinert. 
The pressure loss due to the multiplier has been calibrated and 
does not exceed $0.3$ kbar in the pressure range 
used  in this study, 
so that the accuracy of the pressure in these experiments is better than $\pm 0.15$ kbar.

\section{Spin dynamics in the metallic state}

In weakly correlated elemental metals, $\mathrm{Im}\chi $ is related to
$\rho (E_{F})$, the density of states at the Fermi level, 
and $A_{\mathrm{hf}}$ is $\mathbf{q}$ independent for a 
purely local on-site hyperfine coupling.
Equation (\ref{eq:T1-g}) resumes in that case into the Korringa relation 
\begin{equation}
\frac{1}{T_{1}T}=\frac{\pi k_{B}}{\hbar }\left[ A_{\mathrm{hf}}\rho (E_{F})%
\right] ^{2}.  \label{eq:T1-FL}
\end{equation}

\begin{figure}[t]
\begin{center}
\includegraphics[width=8.5cm]{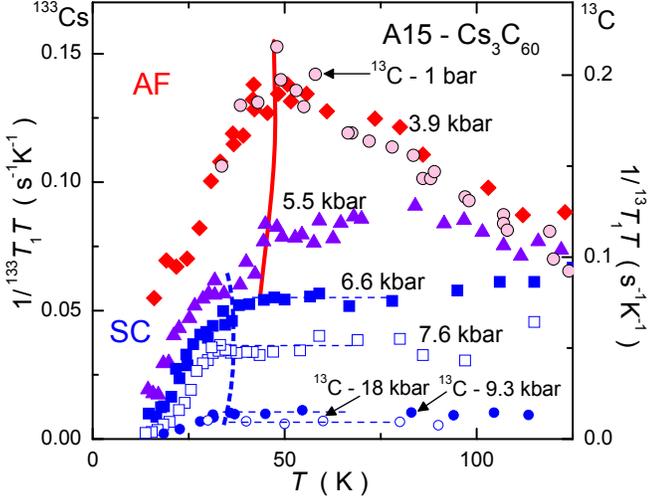}
\end{center}
\caption{(colour on line)In the A15 phase, $(T_{1}T)^{-1}$ has  a large $T$
variation in the paramagnetic insulating state at $p=1$ bar and drops at the
AF transition (full thick red line). At high $p$ a constant Korringa like
behaviour is seen above $T_{c}$, with the expected drop below $T_{c}$
(thick dotted blue line) due to the opening of the the SC gap. Notice the
large increase of the "Korringa" constant (dotted thin line) 
when approaching the $\approx 6.5$
kbar Mott transition. The $^{13}$C data (circles) and the $^{133}$Cs data
have been scaled as discussed in the text.}
\label{T1A15}
\end{figure}

Such a variation is of course not observed in the Mott insulating paramagnetic state.
Indeed there, as can be seen for instance in  
Fig.~\ref{T1A15} for the A15 phase, 
$(T_{1}T)^{-1}$ increases markedly at low $T$ for both $^{13}$C 
and $^{133}$Cs nuclei and drops abruptly at the N\'{e}el
transition at $T_{N}=47$ K. At the highest pressures, well into the metallic
phase, the $(T_{1}T)^{-1}$ data display the constant expected metallic
behaviour. But, when $p$ is decreased towards $p_{c}\approx 6.5$ kbar, the
metallic behaviour is seen only above $T_{c}$ up to about $100$ K and
deviations begin to occur at higher $T$
, as will be much better seen on the fcc phase data.

\begin{figure}[tbp]
\begin{center}
\includegraphics[width=7.5cm]{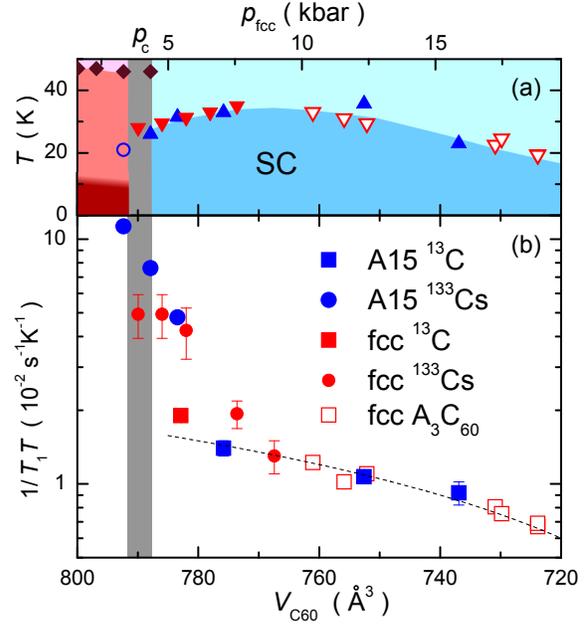}
\end{center}
\caption{(color on line) (a) The phase diagram of 
Fig.~\ref{Common phase diagram} is reproduced here as a reference, 
with the upper pressure scale for the fcc phase. 
(b) Plot versus $V_{C_{60}}$of the $^{13}$C
"Korringa" $(^{13}T_{1}T)^{-1}$ value taken just above $T_{c}$ in the
metallic phases. Here the $^{133}$Cs data displayed have been rescaled for
comparison (see text). For both phases the simple BCS analysis of $T_{c}$,
shown by a dotted line breaks down near $p_{c}$.}
\label{phasediagram}
\end{figure}

Let us consider first here the $^{13}$C NMR data, for which these constant 
$(^{13}T_{1}T)^{-1}$ values taken well above the Mott transition pressure 
$p_{c}$ are plotted then versus $V_{C_{60}}$ in Fig.~\ref{phasediagram}.
There, one can see that for both the A15 and fcc-Cs$_{3}$C$_{60}$ phases the 
$^{13}T_{1}$ data merge with the smooth curve found on the other fcc-$A_{3}$C%
$_{60}$ compounds \cite{maniwa-JPSJ63}.

The corresponding variation of $T_{c}$ versus $V_{C_{60}}$ (or
lattice constant) has been used at length in the past to indicate that a BCS
formalism applies to the fcc-$A_{3}$C$_{60}$ compounds. 
In this framework, $T_{c}$ is given by 
\begin{equation}
k_{B}T_{c}=1.14\hbar \omega _{D}\exp {\left( -1/V\rho (E_{F})\right) }.
\label{eq:bcs}
\end{equation}%
This has led to consider that the Debye frequency $\omega _{D}$ and the
electron-phonon coupling $V$ depend solely on C$_{60}$ molecular properties,
so that a smooth variation of $\rho (E_{F})$ with $V_{C_{60}}$ drives both
variations of $T_{c}$ and $(^{13}T_{1}T)^{-1}$ \cite{gunnarsson-RMP69}. 
Here we observed by r.f. susceptibility measurements that $T_{c}$ goes
through a maximum versus $V_{C_{60}}$ in both phases 
while $(^{13}T_{1}T)^{-1}$ steadily increases. 
This is indicative of a breakdown of Eq.~(\ref{eq:bcs}), as one would
expect a decrease of $\rho (E_{F})$ or of $(^{13}T_{1}T)^{-1}$ when $T_{c}$
goes through the optimum value.

%

To better probe the evolution near the MIT, 
we considered as well $^{133}$Cs data, 
which appear qualitatively similar to those on $^{13}$C,  as shown
in Figs.~\ref{T1A15} and \ref{T1CsFCC}. The raw data are found to display a
large increase of $(^{133}T_{1}T)^{-1}$ with decreasing pressure from the
top of the SC dome towards the Mott transition. This definitely points out
an incidence of electronic correlations. 
\footnote{When magnetic correlations between C$_{60}$ sites enter into
play near the MI transition or near the Mott transition, 
the non local $^{133}$Cs hyperfine coupling $A(\mathbf{q})$ might 
yield differences between the $T$ variations of $^{133}$Cs data 
and of the $^{13}$C $T_{1}$ data, 
as the latter probe directly the on-site magnetic fluctuations. 
However, this may only yield a reduction of the contribution of the AF
fluctuations to $(^{133}T_{1}T)^{-1}$, 
so the actual increase of $\mathrm{Im}\left[ \chi (\mathbf{q},\omega )\right]$ 
when approaching $p_{c}$ could only be larger.} 
The $^{133}$Cs data
for the A15 phase could be scaled with that of $^{13}$C,  
using the ratio $1.4(2)$ obtained from the 
$p=1$ bar data in the paramagnetic state from Fig.~\ref{T1A15}. For the fcc
phase the $^{133}$Cs T' site data plotted in Fig.~\ref{phasediagram} have
been scaled by fixing the $9$ kbar point of Fig.~\ref{T1CsFCC} on the dotted
curve known for the smaller $V_{C_{60}}$ values. 
The scaled $^{133}(T_{1}T)^{-1}$ values in Fig.~\ref{phasediagram}  
display a large
deviation near the Mott transition which confirms that Eqs.~(\ref{eq:T1-FL})
and (\ref{eq:bcs}) do not apply.

\section{Metal Insulator transition}

\begin{figure}[tbp]
\begin{center}
\includegraphics[width=7.5cm]{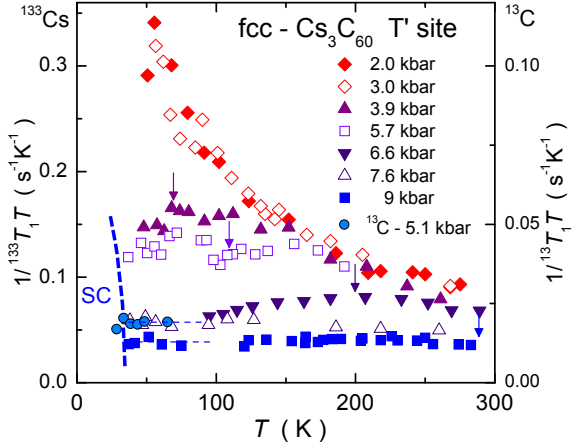}
\end{center}
\caption{(color on line) $(^{133}T_{1}T)^{-1}$ for the T' site of the
fcc-phase. For each pressure above $p_{c}=3.5$ kbar, 
the data have been taken above $T_{c}$ which is delineated by the thick dotted blue line. 
$(T_{1}T)^{-1}$ departs with increasing $T$ from the constant Korringa value 
(thin dotted blue line), goes through a maximum and decreases 
then towards the data obtained in the paramagnetic Mott state. 
The arrows are the positions of $p_{c}(T)$ determined as explained later in
Fig.~\ref{3DFigure} and \ref{Critical}.}
\label{T1CsFCC}
\end{figure}

Let us now consider our measurements versus $T$ for various pressures which
we have performed mostly on $^{133}$Cs NMR in order to benefit from the
phase selectivity allowed by the large difference of spectra between the two
phases on these nuclei (see Ref.~\cite{ihara-PRL104}). 
The data
displayed in Fig.~\ref{T1A15} and Fig.~\ref{T1CsFCC} are qualitatively
similar for both phases. 
However in the following part of the paper 
we shall restrict to the fcc phase on which we could 
take reliable data in a large range of pressures and temperatures. 
There we have measured $T_{1}$ on 
the three Cs sites O, T, T' but did find that at low $T$ the signals of the
O and T sites become so close in frequency that they can hardly be resolved. 
This becomes even impossible below $T_{c}$ due to the loss of spin
susceptibility associated with singlet pairing. So we restricted the data
displayed in Fig.~\ref{T1CsFCC} to that above $T_{c}$ for the T' site which
is better separated from the T site and has a $T_{1}$ close to that of the T
site, which ensures that nuclear-spin cross relaxation effects do not
influence significantly the data. 

For $p>p_{c}$, the $(^{133}T_{1}T)^{-1}$ data differ
markedly at low $T$ from that in the Mott state, go through a maximum, and
then decrease progressively at high temperature 
, as can be seen in Fig.~\ref{T1A15}. The
regime of constant $(^{133}T_{1}T)^{-1}$, which
extends nearly to room $T$ in the higher pressure cases, reduces to smaller 
$T$ ranges when $p$ decreases towards $p_{c}$. 
This maximum in $(T_{1}T)^{-1}$ might bear some
analogy to that seen in underdoped cuprates, and could be
hastily attributed to a pseudogap \cite{alloul-PRL63}, 
or spin gap in the magnetic excitations \cite{yoshimura-JPSJ58}. 
This maximum occurs at a temperature $T^{\ast }$
which increases when $p$ (and $T_{c}$) increase on the superconducting dome. 
Then, this analogy does not hold as $T^{\ast }$ in cuprates decreases with increasing 
doping \cite{alloul-PRL63} towards the optimum $T_{c}$. 

More importantly, one can  notice in  Fig.~\ref{T1CsFCC} 
that the $(^{133}T_{1}T)^{-1}$ data for $p>p_{c}$, after
reaching their maximal values, evolve progressively at high $T$ towards the 
$(^{133}T_{1}T)^{-1}$ data measured below $p_{c}$ in the Mott insulating
paramagnetic state. There the localized electron spins exhibit Curie-Weiss
behaviour and for a dense  local-moment paramagnet, one expects local
fluctuation of the moments dominated by the exchange coupling $J$ with their 
$z$ neighbors, with a high temperature limit 
\begin{equation}
\frac{1}{T_{1}}=\left[ A_{\mathrm{hf}}\right] ^{2}\frac{\sqrt{\pi S(S+1)}}{%
\sqrt{3}\gamma _{e}\mu _{B}k_{B}zJ}.  \label{eq:T1-local}
\end{equation}%
This fits nicely the previous observations in A15  and fcc phases
\cite{ihara-PRL104} for which $(T_{1}T)^{-1}$ scales at high 
$T$ with $T^{-1}$  \cite{jeglic-PRB80, ihara-PRL104}. 
Then, in the fcc phase for pressures between $p_{c}=3.5(5)$ kbar and
$6.6$ kbar, we see in Fig.~\ref{T1CsFCC} that at high $T$ the $(T_{1}T)^{-1}$
data seems to keep initially the same behaviour as that found in the Mott
state. At lower $T$ the excitations are progressively suppressed and
transform into the Korringa like behaviour discussed above in the metallic
state.
\footnote{
This points again at a different situation from that found for
the cuprate pseudogap. 
Indeed in that case the pseudogap is a depression at low $T$ of the DOS 
in a high $T$ metallic state, 
while here the depression  occurs in a Curie paramagnetic state and 
results in a low $T$ metallic state.} 

\begin{figure}[tbp]
\begin{center}
\includegraphics[width=8cm]{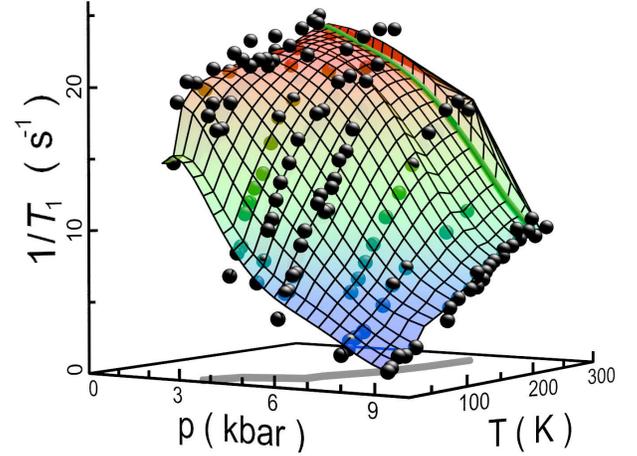}
\end{center}
\caption{(color on line) The $T_{1}^{-1}$ data for the $^{133}$Cs T' site of
fcc-Cs$_{3}$C$_{60}$ reported in Fig.~\ref{T1CsFCC} are shown here
in a 3D plot versus $(p,T)$. The data can be approximated by the smoothed
surface represented as a coloured grid, the black dots being above the grid
while those below appear coloured, being seen by transparency. The
transition $p_{c}(T)$ from the low $p$ insulating state constant 
$T_{1}^{-1}$ above $100$ K to the high $p$ nearly $T$ linear behaviour in
the metallic state is evident. The transition occurs at $p_{c}(T)$,
projected in the ground plane, which shifts towards higher $p_{c}$ with
increasing $T$. A full line at $T=250$ K gives an example permitting to view
the inflexion point in the variation of $1/T_{1}(p)$ at $p_{c}=6.6$ kbar,
which corresponds to the arrow shown in Fig.~\ref{T1CsFCC} for the
corresponding isopressure data.}
\label{3DFigure}
\end{figure}

To monitor differently the evolution of the local moment contributions to 
$T_{1}^{-1}$ with increasing pressure we have therefore plotted the data of
Fig.~\ref{T1CsFCC} in a 3D plot of $T_{1}^{-1}$ versus $(p,T)$ shown in 
Fig.~\ref{3DFigure}. 
One can see as well that at high $T$ the data converges 
for all pressures below $7$ kbar towards the constant value 
found in the ambient pressure insulating state.
Here one can see that the data  below $p_{c}=3.5$ kbar are nearly 
constant at high $T$. 
The decrease detected below $100$ K  has been assigned \cite{ihara-PRL104}
to the incidence of 
interactions between the local moments. This leads to a decrease in 
$(^{133}T_{1})^{-1}$ much weaker in the frustrated magnet fcc phase 
\cite{ihara-PRL104} than in the real gap opening N\'{e}el phase 
of the A15-Cs$_{3}$C$_{60}$.  
In this representation one is then led to view isothermal lines, in which
the  $T_{1}^{-1}$ data, constant at low $p$, in the insulating state drops
with increasing  $p$  towards the metallic behaviour.  Indeed if we look
at the surface which approximates all the data points in 
Fig.~\ref{3DFigure}, 
the drop at $p_{c}$ from the low $p$ to the high $p$ behaviour at constant $T$ and 
from the high $T$ to the low $T$ behaviour at constant $p$ can be seen to
occur at inflexion points $T_{\rm MIT}$ of the fitting grid. Those correspond
in the latter case to the maxima indicated by arrows in Fig.~\ref{T1CsFCC}.
We do not see here  very sharp steps of $T_{1}^{-1}$ as would be
expected for a first order transition. This appears quite natural 
as no special care was taken to see hysteresis effects 
in these isopressure measurements when varying the temperature.
The large set of data obtained here however allows us to locate the transition
pressure $p_{c}(T_{\rm MIT})$ at which such a step like decrease of $T_{1}^{-1}$
is observed at fixed $T=T_{\rm MIT}$.

\begin{figure}[tbp]
\begin{center}
\includegraphics[width=7.5cm]{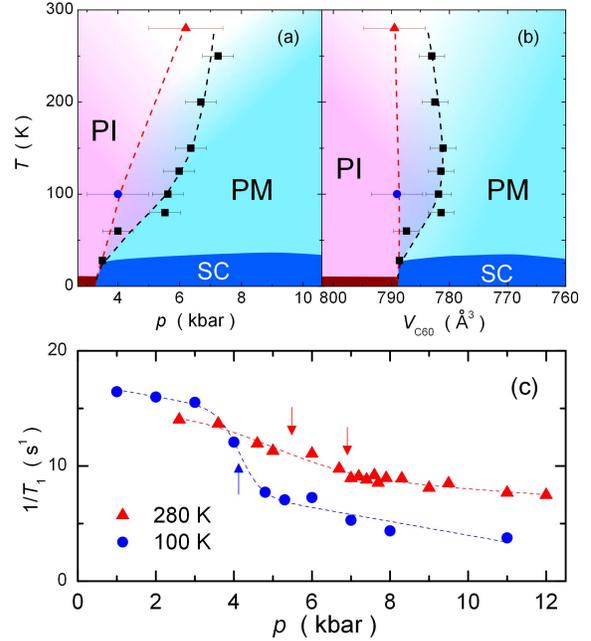}
\end{center}
\caption{(color on line) (a) Variation of $T_{\mathrm{MIT}}$ versus $p_{c}$
obtained from the analysis of the two sets of 
experiments of Fig.~\ref{3DFigure} (squares) 
and of (c) below (circles and triangles). The various
phases of the low $T$ phase diagram of Fig.~\ref{Common phase diagram} 
are reproduced. (b) $T_{\mathrm{MIT}}$ is found nearly independent
of $V_{C_{60}}$ after correcting for lattice volume variations. (c)
Systematic $T_{1}^{-1}(p)$ data taken on $100$ K and $290$ K isotherms show
that the $T_{1}^{-1}$ step at $p_{c}(T)$ decreases markedly at room $T$
(for the arrows: see text) which would be near the critical point for this
fcc phase sample (all dotted lines in the three panels are guides to the
eye).}
\label{Critical}
\end{figure}

This reveals that the transition pressure $p_{c}$ indeed increases with
increasing $T$ so that the $T_{\rm MIT}(p_{c})$ line is not vertical 
in the $(p,T)$ plane, as shown in Fig.~\ref{Critical}(a). 
One should however remind
that these materials are quite compressible and have a high thermal
expansion, the dilatation from 0 to $250$ K corresponding to a $15$ 
 \textperthousand~ 
increase of $V_{C_{60}}$. Therefore the data
plotted in Fig.~\ref{3DFigure} correspond to variable inter-ball distance.
If one takes into account the variation of the lattice constant with $p$ and 
$T$ from Ref.~\cite{ganin-Nature466}, one can as well plot the Mott
transition temperature $T_{\mathrm{MIT}}$ versus $V_{C_{60}}$ as done on
Fig.~\ref{Critical}(b). It is remarkable then to find out that this leads to
a nearly vertical line, which indicates that the transition is indeed
governed by interball distance, that is by the magnitude of the transfer
integral or $U/W$ ratio. 

Let us point out that a decrease of step height with increasing $T$ could be
guessed in the set of data of Fig.~\ref{T1CsFCC} and \ref{3DFigure}. However
these data being only taken at that time for a limited set of pressure
values we could not quantify the step height values accurately. To get
better data on the evolution of $T_{1}^{-1}$ at the transition we have then
performed a more systematic study of its variation with $p$ first at a fixed 
$T=290$ K and then at $T=100$ K. Here we do find now as seen in 
Fig.~\ref{Critical}(c) that at $100$ K the step of $T_{1}^{-1}$ 
is quite visible and
occurs at $4$ kbar near the $3.5$ kbar value obtained from the diamagnetic
susceptibility data \cite{ihara-PRL104}. At $290$ K the variation of 
$T_{1}^{-1}$ is rather smooth so that the transition is more difficult to
locate. The lower $p$  arrow in Fig.~\ref{Critical}(c), would correspond to
a widened transition, while the second one would correspond to a crossover
resuming in a  kink at $7$ kbar.  Both cases would indicate that one
approaches a critical point at room $T$, as in a liquid-gas transition 
\footnote{The data of Fig.~\ref{Critical}(c) has been taken six month later 
than Fig.~\ref{T1CsFCC} and \ref{3DFigure}. 
After curing such a long time at room $T$ in the pressure cell 
a slight decrease of $T_{1}^{-1}$values as compared to those of 
Fig.~\ref{3DFigure} has been observed together with a small 
shift of the transition towards lower pressures displayed in Fig.~\ref{Critical}a. 
Such an evolution might be assigned to a small increase 
of sample metallicity, that is an increase of sample density 
at ambient pressure.}, 
in  analogy to what has been seen in other Mott transitions
i.e. in organic compounds \cite{kagawa-NP5}.

Finally, as had been suggested for the metallic state of the
dense $A_{3}$C$_{60}$ compounds \cite{brouet-PRB66}, excitations from the 
$S=1/2$ to the $S=3/2$ Jahn Teller molecular state are expected at
sufficiently high $T$. Those should be seen already in the insulating state
and should appear as a high $T$ increase of $T_{1}{}^{-1}$, as expected
from Eq.~(\ref{eq:T1-local}) for $S=3/2$. 
As seen in Fig.~\ref{3DFigure} we did not detect
any such increase at $p=1$ bar for fcc-Cs$_{3}$C$_{60}$, so that these
excitations would only be detectable above room $T$. Capone \textit{et al.} 
predicted that the energy distance between the two molecular states should
decrease in the metallic state below the optimal 
$T_{c}$, yielding a situation reminiscent to that of a pseudogap. So far, the
variation with pressure of the transition line does not allow us here to
probe these excitations, 
which apparently might only occur above room $T$, 
even in the metallic states.

\section{Discussion}

Here we have performed $^{133}$Cs and $^{13}$C $T_{1}$ measurements under
pressure which bring quite novel insight on SC near the 
Mott transition at $p_{c}$ in the fcc-$A_{3}$C$_{60}$ compounds, 
and on the Mott transition itself in the fcc phase of Cs$_{3}$C$_{60}$. 
We have shown that, just above $T_{c}$ in the metallic state, 
$(T_{1}T)^{-1}$deviates in both phases near $p_{c}$ from the trend 
observed for the dense $A_{3}$C$_{60}$ compounds. The
large observed increase of $(T_{1}T)^{-1}$ indicates that spin fluctuations
in the metallic state become prominent before the system switches into the
Mott state.

Capone \textit{et al.} \cite{capone-Science296, capone-RMP81} were
suggesting a moderate increase of the spin susceptibility when approaching
the MIT, which would lead to deviations from the smooth $T_{c}$ versus DOS
BCS prediction. The significant increase of $(T_{1}T)^{-1}$ found here is 
similarly an
evidence that the $T_{c}$ variation deviates then from pure BCS but
it is also a signature of strong magnetic correlations near the transition.
Calculations of the dynamic susceptibility within their theoretical scheme
should help to decide then whether the large $T_{c}$ values in  
$A_{3}$C$_{60}$ compounds does indeed result from a  
fundamental cooperation between
spin fluctuations and electron phonon interactions.  

We have shown here that the variations of $(T_{1}T)^{-1}$ versus $T$ at
fixed $p$ although analogous to those expected for a pseudogap, mainly
monitor the shift of the transition pressure with temperature due to lattice
expansion. The large $T$ dependence of the transition pressure $p_{c}$ found
here is somewhat similar to that seen in 2D-organics. 
In the present experiments, the 
"liquid-gas like" critical point of the phase diagram is located near room 
$T$. While in the 2D organics the Mott transition might be influenced by the
pressure induced increase of interplane coupling, in the present 3D system
we are probing a simple variation of $U/W$ through the reduction of lattice
parameter. Therefore, in the $(p,T)$ range probed here the qualitative
behaviour detected appears quite similar to those expected for a single
orbital Mott transition \cite{georges-RMP68}, although the $S=1/2$ ground
sate is orbitally degenerate.

To conclude, the local probe studies of Cs$_{3}$C$_{60}$ provided here 
reveal that it is certainly a unique system in which 3D multiorbital
behaviour coexists with singlet $s$ wave fully gapped SC and spin
fluctuations in the metallic state near the MIT. The fact that we could not
evidence here the excitations between the low spin and high spin molecular
states calls for ab initio determination of their energy difference.
Measurements of other thermodynamic and spectral properties, possibly on
larger $T$ ranges are required to complete the experimental insight on this
Mott transition.

\acknowledgements
We thank V.~Brouet, M.~Capone, M.~Fabrizio, F.~Rullier-Albenque and 
E.~Tosatti for stimulating exchanges
about these experimental results and careful reading of the manuscript. 
Y.~I. acknowledges JSPS Postdoctral Fellowships for Research Abroad. 


\begin{thebibliography}{10}
\expandafter\ifx\csname url\endcsname\relax\def\url#1{\texttt{#1}}\fi

\bibitem{takabayashi-Science323}
\Name{Takabayashi Y., Ganin A.~Y., Jegli{\v{c}} P., Ar{\v{c}}on D., Takano T.,
  Iwasa Y., Ohishi Y., Takata M., Takeshita N., Prassides K. \and Rosseinsky
  M.~J.} \REVIEW{Science }{323}{2009}{1585}.

\bibitem{ihara-PRL104}
\Name{Ihara Y., Alloul H., Wzietek P., Pontiroli D., Mazzani M. \and Ricc\`{o}
  M.} \REVIEW{Phys.~Rev.~Lett. }{104}{2010}{256402}.

\bibitem{ganin-Nature466}
\Name{Ganin A.~Y., Takabayashi Y., Jegli{\v{c}} P., Ar{\v{c}}on D.,
  Poto{\v{c}}nik A., Baker P.~J., Ohishi Y., McDonald M.~T., Tzirakis M.~D.,
  McLennan A., Darling G.~R., Takata M., Rosseinsky M.~J. \and Prassides K.}
  \REVIEW{Nature }{466}{2010}{221}.

\bibitem{lee-RMP78}
\Name{Lee P.~A., Nagaosa N. \and Wen X.-G.} \REVIEW{Rev.~Mod.~Phys.
  }{78}{2006}{17}.

\bibitem{alloul-EPL91}
\Name{Alloul H., Rullier-Albenque R., Vignolle B., Colson D. \and Forget A.}
  \REVIEW{EPL }{91}{2010}{37005}.

\bibitem{gunnarsson-RMP69}
\Name{Gunnarsson O.} \REVIEW{Rev.~Mod.~Phys. }{69}{1997}{575}.

\bibitem{tsuei-PRL73}
\Name{Tsuei C.~C., Kirtley J.~R., Chi C.~C., Yu-Jahnes L.~S., Gupta A., Shaw
  T., Sun J.~Z. \and Ketchen M.~B.} \REVIEW{Phys.~Rev.~Lett. }{73}{1994}{593}.

\bibitem{lefebvre-PRL85}
\Name{Lefebvre S., Wzietek P., Brown S., Bourbonnais C., J\'{e}rome D.,
  M\'{e}zi\`{e}re C., Fourmigu\'{e} M. \and Batail P.} \REVIEW{Phys.~Rev.~Lett.
  }{85}{2000}{5420}.

\bibitem{capone-PRB62}
\Name{Capone M., Fabrizio M., Giannozzi P. \and Tosatti E.}
  \REVIEW{Phys.~Rev.~B }{62}{2000}{7619}.

\bibitem{capone-Science296}
\Name{Capone M., Fabrizio M., Castellani C. \and Tosatti E.} \REVIEW{Science
  }{296}{2002}{2364}.

\bibitem{capone-RMP81}
\Name{Capone M., Fabrizio M., Castellani C. \and Tosatti E.}
  \REVIEW{Rev.~Mod.~Phys. }{81}{2009}{943}.

\bibitem{kagawa-NP5}
\Name{Kagawa F., Miyagawa K. \and Kanoda K.} \REVIEW{Nature Physics
  }{5}{2009}{880}.

\bibitem{pennington-RMP68}
\Name{Pennington C.~H. \and Stenger V.~A.} \REVIEW{Rev.~Mod.~Phys.
  }{68}{1996}{855}.

\bibitem{brouet-PRL82}
\Name{Brouet V., Alloul H., Qu{\'{e}}r{\'{e}} F., Baumgartner G. \and
  Forr{\'{o}} L.} \REVIEW{Phy.~Rev.~Lett. }{82}{1999}{2131}.

\bibitem{brouet-PRB66}
\Name{Brouet V., Alloul H., Garaj S. \and Ferr{\'{o}} L.} \REVIEW{Phys.~Rev.~B
  }{66}{2002}{155122}.

\bibitem{maniwa-JPSJ63}
\Name{Maniwa Y., Saito T., Ohi A., Mizoguchi K., Kume K., Kikuchi K., Ikemoto
  I., Suzuki S., Achiba Y., Kosaka M., Tanigaki K. \and Ebbesen T.~W.}
  \REVIEW{J.~Phys.~Soc.~Jpn. }{63}{1994}{1139}.

\bibitem{alloul-PRL63}
\Name{Alloul H., Ohno T. \and Mendels P.} \REVIEW{Phys.~Rev.~Lett.
  }{63}{1989}{1700}.

\bibitem{yoshimura-JPSJ58}
\Name{Yoshimura K., Imai T., Shimizu T., Ueda Y., Kosuge K. \and Yasuoka H.}
  \REVIEW{J.~Phys.~Soc.~Jpn }{58}{1989}{3057}.

\bibitem{jeglic-PRB80}
\Name{Jegli{\v{c}} P., Ar{\v{c}}on D., Poto{\v{c}}nik A., Ganin A.~Y.,
  Takabayashi T., Rosseinsky M.~J. \and Prassides K.} \REVIEW{Phys.~Rev.~B
  }{80}{2009}{195424}.

\bibitem{georges-RMP68}
\Name{Georges A., Kotliar G., Krauth W. \and Rozenberg M.~J.}
  \REVIEW{Rev.~Mod.~Phys. }{68}{1996}{13}.

\end{thebibliography}

\end{document}